# Unraveling the Conglomeratic Nature of Methanol Clusters Adsorbed on Graphene Surfaces. Insights from Molecular Dynamics Simulations†

Juan García de la Concepción,*[a] Izaskun Jiménez-Serra,[b] Ibon Alkorta,[c] José Elguero[c] and Pedro Cintas*[a]

The expression of chirality in adsorption phenomena constitutes an important topic, not only relevant to asymmetric transformations involving solid surfaces, but also to the emergence of homochirality in both terrestrial and extraterrestrial scenarios. Methanol (MeOH) aggregation on graphite/graphene, one of the most idealized adsorbate-adsorbent systems, lead to islands of cyclic clusters of different sizes (*Nano. Lett.*, 2016, **16**, 3142-3147). Here, we show that this aggregation occurs enantioselectively affording 2D conglomerates depending on the size of clusters, in close analogy to a Pasteurian racemate. Homochiral sequences are held together by hydrogen bonding and other non-covalent interactions, whose absolute configurations can be appropriately specified. A discussion involving the dichotomy between 2D racemates and conglomerates, is offered as well. In addition, the present simulations showcase a broad range of acyclic and cyclic structures, even if some discrete rings are the dominant species, in agreement with previous experimental data and theoretical modeling. Our results indicate that MeOH clusters show binding energies close to the experimental values, remaining intact at temperatures as high as 120 K and up to 150 K.

## 1 Introduction

In non-idealized environments, where any chemical system is dynamic even at absolute zero, racemic mixtures of chiral molecules tend to aggregate into heterochiral, rather than homochiral domains. On the other hand, in a static idealized environment, achiral molecules that interact to form 3D clusters can give rise to chiral aggregations. A particularly representative case is the water molecule, which, despite being completely achiral, can form a three-membered cluster in the gas phase that lacks a plane of symmetry.[1] The situation of 2D chirality has remained unclear, although a common hypothesis is that homochiral packing is preferred in surfaces.[2–7] Chirality in two dimensions, either on surfaces or at interfaces, can be categorized according to the type of adsorbate, i.e. inherently chiral, prochiral, conformationally chiral, and achiral.[8] Prochiral adsorbates, which lack chirality in 3D as they have mirror images which can be superimposed, can however adopt non-superimposable configurations by being constrained to 2D upon adsorption.

As expected for other polar molecules, MeOH is prone to self-assembly through H-bonding, and in fact MeOH trimers already exhibit some cooperativity upon binding as evidenced by topological changes in bond distances.[9] Crystalline MeOH polymorphs, generated at high pressure, show likewise supramolecular zig-zag structures between 30-150 K.[10,11] Both theory and experiments indicate that liquid MeOH consists of H-bonded rings and acyclic chains, where hexamers and octamers were identified as the dominant species.[12] DFT studies of MeOH clusters, showed the higher stability of rings, especially five- and six-membered, relative to structures of the same number of MeOH molecules where one or more lie outside the ring, and acyclic frameworks are even less stable.[13]

Concerning the adsorbate-adsorbent interaction, the MeOH-graphite/graphene system is the most extensively studied and idealized, even being chosen as an appropriate model for benchmarking more complex systems. Given the apolar character of the adsorbent, MeOH tends to orient the OH group slightly pointing toward the surface.[14,15] When more MeOH molecules are present, they aggregate into clusters, stabilized by intermolecular H-bonding, which rapidly diffuse along the surface to interact with other clusters or creating new ones at low surface coverages. MeOH clusters form small islands when absorbed on highly-oriented pyrolytic graphite (HOPG),[16] a structural feature employed as starting point to determine the adsorption energy of the first MeOH monolayer on the carbonaceous surface.[17,18]

Whether MeOH molecules break 2D symmetry upon adsorp-tion has not yet been thoroughly analyzed in detail. Probably, the first genuine evidence was reported in 2011 by Sykes and associates, who obtained low-temperature scanning tunneling microscopy (STM) images of enantiopure MeOH hexamers adsorbed on Au(111).[19] It is clear that MeOH hexamers adopt a homochiral sequence and the authors rightly identified that symmetry breaking takes place as binding to the Au surface renders methanol chiral. Notably, the MeOH hexamers are essentially flat and the H-bonds between each monomer are almost linear. While rings other than hexamers were not observed, less stable zig-zag chain structures also occur at higher MeOH coverages. However, in such acyclic arrangements each methanol molecule adopts a distinct configuration.

[a] *Departamento de Química Orgánica e Inorgánica, Facultad de Ciencias, and IACYS-Green Chemistry and Sustainable Development Unit, Universidad de Extremadura, 06006 Badajoz, Spain; E-mail: jugarco@unex.es; pecintas@unex.es*
[b] *Departamento de Astrofísica, Centro de Astrobiología (CSIC-INTA), Carretera de Torrejón-Ajalvir, Km 4, Torrejón de Ardoz, 28850 Madrid, Spain.*
[c] *Instituto de Química Médica (IQM-CSIC), Juan de la Cierva 3, 28006 Madrid, Spain.*
† Supplementary Information available: [Further details about the geometries obtained during the MD simulations and stability of the clusters]. See DOI: 10.1039/cXCP00000x/

Thorough analyses by Nguyen at al. in 2011 and Zangi and Roccatano in 2016 using molecular dynamics (MD) simulations of MeOH molecules over graphite and confined between two, or adsorbed on, graphene sheets, respectively, evidenced the formation of monolayers involving H-bonded MeOH chains, which undergo string-to-ring transitions depending on the temperature and 2D density.[18,20] Zangi and Roccatano showed the preferred formation of rings at lower densities appears to be controlled by entropic effects, as strings with only one H-bond at their edges will have an extra enthalpic penalty. Without explicitly alluding to chirality, some snapshots from the MD simulations show mirror-image arrangements of MeOH rings.

With the above premises, the objectives of this study are twofold. Firstly, we revisit the diffusion of MeOH molecules on a graphene surface at various MeOH concentrations and a maximum temperature of 200 K to elucidate more comprehensively the probability of forming cycles of different sizes at temperatures close to experimental conditions and their stability over time at temperatures below the desorption threshold. Additionally, we evaluate the binding energy of MeOH molecules forming these clusters. Next up, this analysis sheds light into the enantiospecific appearance and the conglomeratic nature of 2D MeOH conglomerate clusters as they assemble along the graphene monolayer, its origin and unambiguous oxygen configurations.

# 2 Computational details

## 2.1 Cluster Preparation

To build the clusters used for molecular dynamics simulations, a graphite sheet composed of 198 carbon atoms, forming a layer measuring 36.3 x 21.7 Å was constructed, where the edge carbon atoms were replaced by hydrogen atoms. This structure, along with the methanol molecule, was optimized without constraints using the GFN-FF force field.[21] Subsequently, the methanol molecule was positioned on the graphite sheet by employing the automated interaction site screening (aISS) submodule of the xTB program package.[22,23] The most stable structure was selected from a screening of 20 different positions, which were optimized using the aforementioned method. In the resulting structure, the methanol molecule was located in the center of the graphite sheet with an arrangement almost identical to that obtained from calculations using the generalized gradient approximation of the PBE functional.[15] This final structure was used in the subsequent aISS calculation. In this case, a directed docking calculation was performed, applying an attractive potential towards the methanol molecule and all carbon atoms located within 4 Å of the methanol, using a scaling factor of 0.01 for this attractive potential. This process was repeated to obtain clusters of 4, 5, 6, 7, 8, 9, 10, and 20 methanol molecules on the graphite sheet.

## 2.2 Molecular Dynamics Simulations

Each of the 8 clusters obtained following the previously described methodology was used as input for molecular dynamics calculations. Simulations were performed in the NVT ensemble. The dynamics were propagated for 1000 picoseconds (ps) for each cluster at 200 K using a Berendsen thermostat with a time step of 0.2 fs in order to perform simulations in the canonical NVT ensemble. The mass of the hydrogen atoms was adjusted to 1 atomic unit (au). To prevent the "evaporation" of methanol molecules, the system was encapsulated in an ellipse with radii of 20.1, 13.1, and 4.8 Å using a logfermi potential scaled at a temperature of 500 K. In some simulations, the graphite sheet folded, and calculations were then performed by constraining the graphite sheet without exact fixing. The atoms belonging to the graphite sheet (including the hydrogen atoms of the edges) were subjected to a force constant potential of 0.5 to allow slight movement, thus preventing the folding of the sheet. After running the molecular dynamics at 200 K, structures were extracted every 10 ps and used as input for running molecular dynamics under the same conditions, albeit at a temperature of 10 K and propagated during 10 ps. The 100 final geometries after these "freezing" dynamics were optimized using the same method, allowing for a change in total energy of $1 \times 10^{-6}$ $E_{conv}/E_h$ and a change in gradient norm of $8 \times 10^{-4}$ $G_{conv}/E_h \alpha^{-1}$.

### 2.2.1 Stability analyses

The stability of the most representative MeOH clusters (see below) was evaluated by propagating molecular dynamics at 150 K for 100 ps from fully optimized structures. For the pre-optimization prior to molecular dynamics, we started from the most stable structure of all those obtained with the methodology described in section 2.2.

In the case of the seven-membered cluster, hydrogen bonds breaking was observed during the molecular dynamics. Considering that 150 K is a high temperature, close to the melting point of MeOH, the same molecular dynamics was propagated for 100 ps at 120 K.

## 2.3 Binding energies

Binding energies (BE) of methanol in the clusters were calculated as:

$$E_{BE} = E_{clu} - (E_{clu-1} + E_{MeOH}) \quad (1)$$

where $E_{clu-1}$ is the optimized geometry of the cluster minus one methanol molecule. $E_{MeOH}$ is the energy of the optimized methanol molecule and $E_{clu}$ is the energy of the optimized whole cluster. Binding energies were refined by including the contri-bution of the zero point vibrational energy (ZPE) and calculated as:

$$\Delta ZPE_{BE} = ZPE_{clu} - (ZPE_{clu-1} + ZPE_{MeOH}) \quad (2)$$

The optimizations were carried out with the semiempirical tight-binding GFN2-xTB method, [24] setting a threshold in the total energy and gradient change of $5\times10^{-8}\ E_{conv}/E_h$ and $5\times10^{-5}\ G_{conv}/E_h\alpha^{-1}$, respectively. Vibrational frequencies were obtained within the harmonic approximation with the same method, and obtaining no imaginary frequencies.

## 3 Results

### 3.1 Surface MeOH clustering

With the preliminary goal of determining the energy distribution of MeOH molecules adsorbed on a graphene sheet, the geometries obtained for all clusters were sorted and analyzed in ascending order of energy. Fig.1 shows histograms of the structure counting as a function of the interaction energy in kcal mol$^{-1}$ between the MeOH cluster and the graphene sheet, as well as the 3D-structures of the most stable configurations. The relative energy refers to the energy of a given MeOH cluster on the sheet with respect to the sum of the energies of the MeOH molecules and the surface. Accordingly, that energy cannot be strictly compared to the experimental values obtained for MeOH binding energies (BE), but they reflect the most stable arrangement along a single simulation.

The MD for each set of MeOH molecules exhibited very different behaviors. Therefore, to obtain representative distributions of the structures extracted from the simulations, the bins had to be spaced differently. Even so, all simulations indicate that the most stable structures are rings consisting of H-bonded MeOH molecules acting both as donors and acceptors. In this arrangement, the second lone electron pair on the oxygen atoms points toward the graphene sheet. This type of cyclic non-covalent arrangement was obtained in 4 out of 100 structures for the simulation involving four MeOH molecules (Fig. 1a), 28 for the one with five molecules (Fig.1b), 40 for six (Fig.1c), 56 for seven (Fig.1d) and 17 for eight (Fig.1e). On the other hand, the simulation with nine MeOH molecules resulted in the most stable aggregations being cycles of seven and eight members, with one or two MeOH molecules outside the cycle. The only simulations where zig-zag type structures were observed involved ten MeOH molecules together with the surface saturated by twenty MeOH molecules (Fig. S1). This behaviour has been observed in previous MD simulations as well (see below). [20]

### 3.2 CIP-based oxygen configurations

While MeOH lacks enantiomorphism in 3D, the obvious 2D chirality arises from the non-covalent interactions, as the two lone pairs on the oxygen atom are prochiral and their binding affords a stereogenic oxygen with a distorted tetrahedral geometry bearing four different substituents. Overall, this arrangement forces all the hydrogens in the cluster to orient towards the same direction, giving the structure a unidirectional circular orientation that confers a specific configuration to each MeOH molecule (see optimized geometries of MeOH clusters in Figure 1). MD simulations evidence unambiguously that MeOH clusters self-assembled on the first monolayer are homochiral regardless of their size. Fig.2 shows the moment when the six-membered homochiral cluster is formed during a molecular dynamic simulation conducted at 200 K. In that picture, the colored lines represent the variation of hydrogen bond distances, while the black line depicts a smooth of the electronic energy throughout the simulation. When the simulation captures the formation of all hydrogen bonds there is a drop in electronic energy. That event is highlighted with a vertical pink line at approximately 250 ps.

At this stage, however, to denote the configuration in terms of "orientation" or right-handed/left-handed is confounding. Such notations are generally appropriate for chiral lattices with periodical arrangements, in which right- and left-handed (whether overlayer or substrate) imply the use of right- and left-handed coordination, respectively. [2] Raval and coworkers have shown that the enantiospecific interaction of homochiral molecules with a surface may also create an adsorption footprint, i.e. the bonding connectivity to the surface, if this process leads to the absence of reflection symmetry. [4] A typical case where the orientational effect makes sense is provided by the honeycomb lattice of highly oriented pyrolytic graphite, for which symmetry directions can be identified. A given substrate could then be adsorbed with an angle relative to a unit-cell vector of the graphite, thereby showing the orientation of the monolayer. [25]

In the present case, the graphene sheet does not induce any asymmetry during the deposition of an adsorbate such as MeOH, and the configuration of every homochiral cluster can be assigned with confidence according to the standard CIP (Cahn-Ingold-Prelog) rules. Since the atomic stereogenicity now involves both covalent and non-covalent interactions, the unequivocal assignment using a hierarchical system requires an additional priority rule, as already described. [26] The presence of two non-covalent interactions involving lone pairs (LP···A vs LP···B) follows the general criterion that higher atomic number precedes lower. Thus, as an example shown in Fig.3 left, the substituents around the oxygen atom will have the following order of priority: (i) methyl group, ii) hydrogen atom, i. e., the donor H-bond interaction with an adjacent MeOH molecule, (iii) the interaction of methanol's HOMO with the (carbon) surface, and (iv) the acceptor H-bond interaction with the HOMO-1 orbital.

This assignment ensures that all the oxygen atoms in four-, five-, and six-membered clusters have the same configuration. When the hydrogen-bond donor is oriented in a clockwise direction, all oxygen atoms exhibit the $S$ configuration. On the other hand, the seven- and eight-membered cycles are not enantiomerically pure mixtures because the methyl group of the methanol molecule, which points inward toward the cycle, adopts a different configuration (see Fig.3, right).

### 3.3 Binding energies

Binding energies were computed for the most stable clusters mentioned above, with the sole exception of the twenty MeOH aggregate, since the most representative structures involve the for-mation of seven- and eight-membered rings. We chose the most stable clusters for obtaining BE since at the low temperature conditions where experiments are conducted, the most abundant arrangements should be highly populated. Fig.4 plots the BE from four- to ten-membered methanol-containing rings. The results obtained for the corresponding cycles of seven, eight, nine, and ten MeOH molecules were derived from an average of the individual BE of each MeOH molecule in the cycle, because none is conformationally equivalent. The deviation observed for such rings lies in the fact that BE depend on the cluster structure with subtraction of one MeOH molecule. These optimized structures differ significantly depending on which molecule of the cluster is removed, skewing the average BE. Even with the manual modification of such MeOH clusters by bringing all the molecules closer together, the resulting geometry was almost identical to that obtained without altering the original structure.

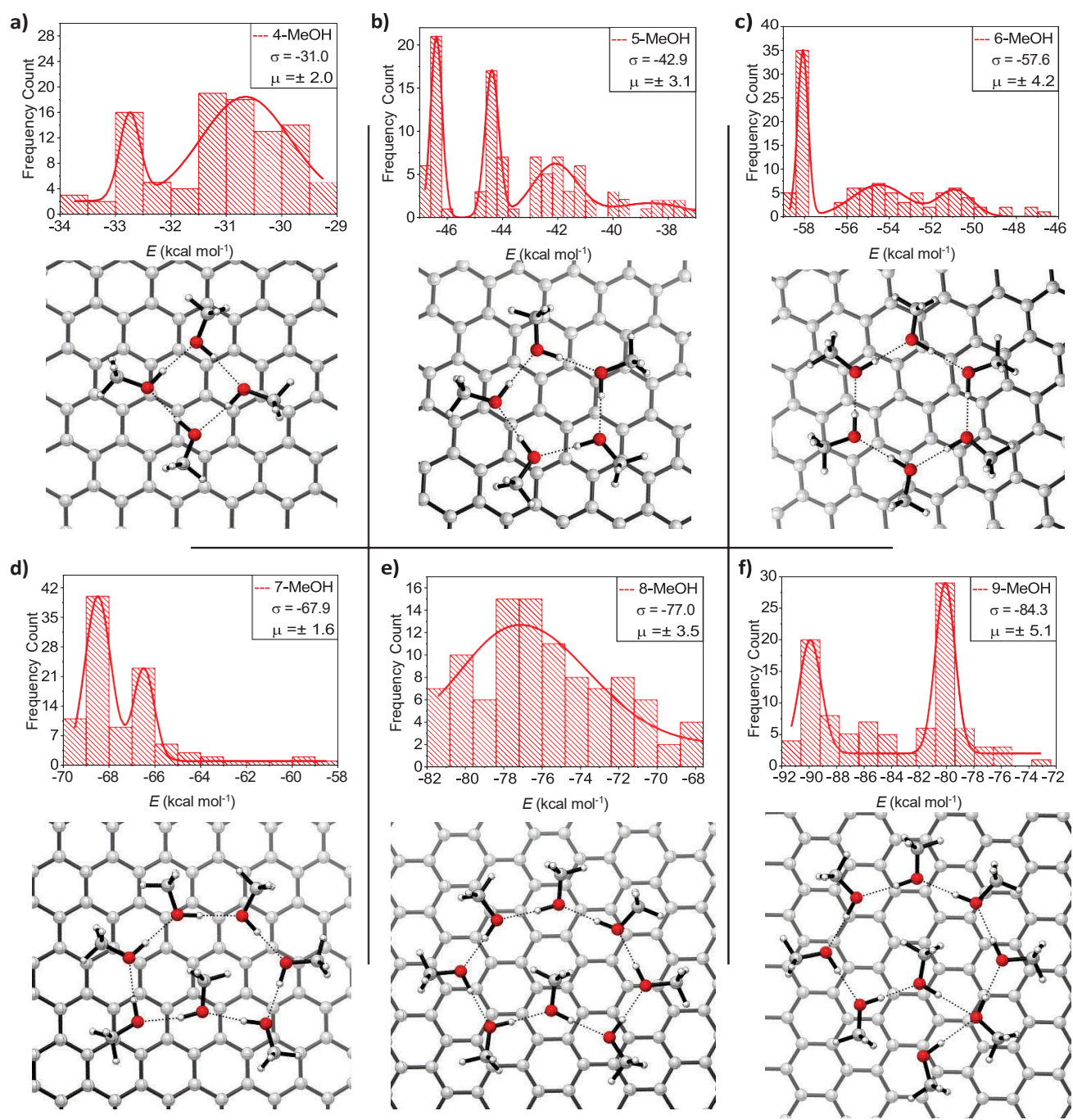

Fig. 1 Distributions of relative energies $\Delta E$ calculated for cluster sizes of 4 (a), 5 (b), 6 (c), 7 (d), 8 (e), and 9 (f) MeOH molecules. The mean ($\mu$) and standard deviation ($\sigma$), both in kcal mol$^{-1}$, for each distribution are shown in the plot legends. Solid red lines are probability density functions fitted to the BE distributions. Different spaced bins were used for each cluster owing to their different behavior. Distributions consist of 100 samples, each calculated with GFN-FF.

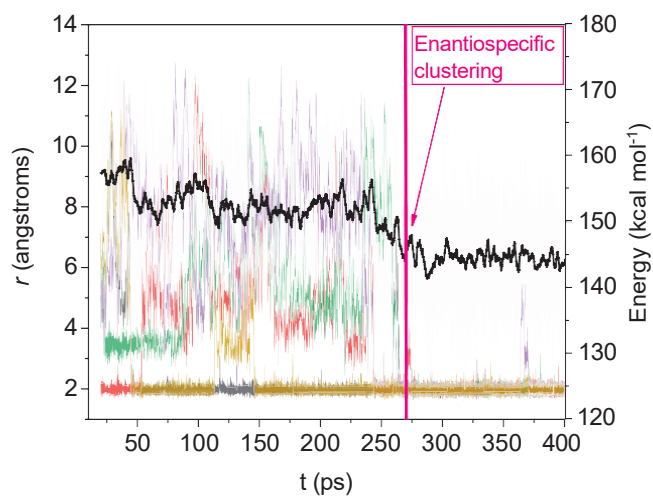

Fig. 2 Variation of H-bonding in six-membered MeOH cluster along a MD simulation at 200 K from randomly oriented MeOH molecules over a graphene sheet during 400 ps. The transparent grey line is the relative electronic energry with respect to the optimized geometry used as starting point. The solid black line corresponds to a Savitzky-Golay smooth with a 100-point window. The vertical pink line indicates the appearance of the homochiral aggregate

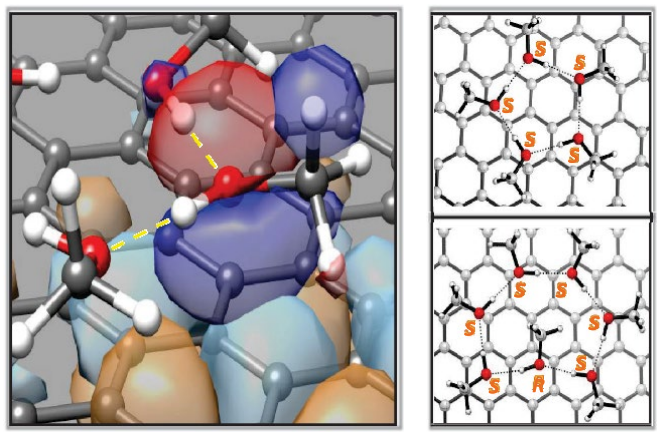

Fig. 3 Left: Zoom-in onto a MeOH molecule in the five-membered ring cluster showing an $S$ configuration. The yellow dashed lines represent the donor and acceptor H-bonds. The interaction of MeOH with the graphene sheet is represented by the overlap of one phase of the methanol's HOMO orbital (dark blue) with the same phase of a $\pi^*$ orbital of a graphene carbon (light blue). Right: Homochiral cyclic cluster of five MeOH molecules and non-homochiral cluster of five MeOH molecules denoting the corresponding configurations at oxygen. It should be noted that the five-membered cluster shows the same configurations as those of four- and six-membered MeOH clusters. The same applies to the comparison between the seven- and eight-membered MeOH clusters

Binding energies alone cannot provide a satisfactory rationale for the MeOH desorption process in view of the large oscillations observed. Despite the observed energy scattering, these re- sults suggest an increasing trend in BE as the number of MeOH molecules included in the cluster increases. The quadratic function depicted in Fig.4 (blue line) seems to provide a smoother fit, although it significantly deviates from the BE obtained for the nine-membered cycle, which should not be ruled out. On the other hand, the logarithmic function fits most of the data better. It is important to note that less stable clusters, after the desorption of one MeOH molecule, lead to higher BE. Therefore, an asymptotic behavior at higher MeOH concentrations seems to better fit the global process under study, where the clusters formed after the desorption of one molecule should rearrange to form stable six-, seven-, or eight-membered rings, and the loss of MeOH will not be detrimental to rearrangements involving many molecules. As shown in Fig.4, the BE exhibits an asymptotic behavior as the number of MeOH molecules increases. This trend is consistent over a broad BE range that qualitatively aligns with experimental data for the first layer of MeOH adsorbed on graphene, where the MeOH density is gradually increased at 105 K.[16]

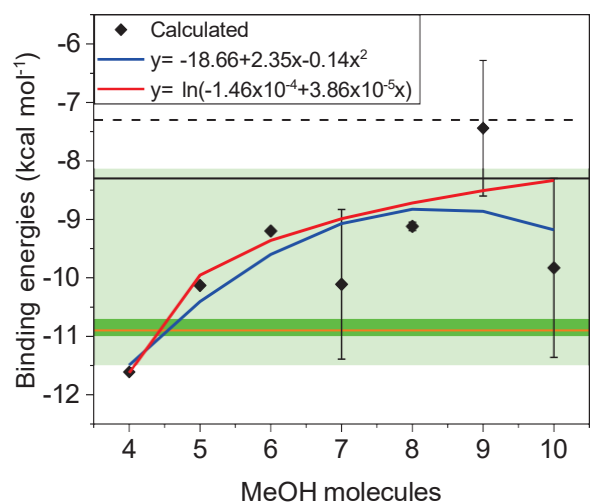

Fig. 4 Calculated BE (ZPE corrected), expressed in kcal mol$^{-1}$, of clusters containing 4-10 MeOH molecules using GFN2-xTB (black squares).[24] Logarithmic (red line) and parabolic (blue line) fittings of the calculated BE. Pale green band corresponds to the BE range obtained for a MeOH monolayer with increasing exposure on a graphite surface at 105 K.[16] Darker green band is the BE range of a submonolayer of MeOH over a graphite surface at 90 K.[17] For comparative purposes, the horizontal orange line is the BE of MeOH in the first monolayer on graphene-covered Pt(111).[27] Horizontal solid and dashed black lines correspond to the BE computed for clusters containing five and three MeOH molecules at DFT level, respectively.[14]

### 3.4 Cluster stability analysis

Finally, we explore the stability of MeOH clusters to assess how long these structures remain over time on the graphene surface. To this end, MD started from completely optimized geometries. Such simulations were conducted for clusters containing from four to seven MeOH molecules, as they are expected to be the most abundant in the first monolayer, as already indicated. Larger clusters (from eight to ten molecules) were ruled out in view of the ten-MeOH cluster yielding seven-membered rings as preva-lent structures. Likewise, no nine-membered rings were detected for the simulation with nine molecules, but rather eight- and seven-membered rings with external MeOH molecules. The productivity of eight-membered cycles is very low when simulating the oligomerization of eight MeOH molecules. Simulations were conducted at 150 K and we discarded any evaluation at higher temperatures as MeOH begins to evaporate at 175 K [16] and becomes crystalline before reaching such temperatures. This observation is consistent with simulations at 200 K, which show both ring-forming and ring-breaking reactions, leading to different arrangements of MeOH on the surface.

The results indicate that only five- and six-membered rings remain intact throughout the entire simulation. On the other hand, four- and seven-membered rings are unstable under these conditions (see Figs. S2, S3, S4 and Fig.5). Fig.5 shows the rearrangement that the seven-membered cluster undergoes during the simulation, which is the most complex of all. In any case, one cannot conclude that four- and seven-membered rings will be absent on the first submonolayer of graphite/graphene, as 150 K is a relatively high temperature for MeOH, close to its melting point. In fact, upon performing another MD simulation of the seven-membered ring at 120 K, the least stable cluster remained unchanged for the entire 100 ps. Therefore, at the typical working temperatures for this system, all mentioned cycles should be present, albeit favoring five- and six-membered MeOH clusters.

## 4 Discussion

### 4.1 Comparison with previous simulations

As already mentioned in Sections 1 and 3.1, Nguyen et al. assessed the behavior of methanol molecules over graphite using Lennard-Jones potentials and fixed partial charges.[18] Later, Zangi and Roccatano carried out a similar study with methanol molecules over and between graphene sheets using the OPLS force field.[20] Both obtained concordant results to those presented in this work. At high densities, they observed that methanol molecules predominantly form hydrogen-bonded linear chains adopting a zig-zag pattern. A trend that we observed when the 36.3 x 21.7 Å graphene sheet supports ten and twenty methanol molecules, namely with an almost and completely saturated surface, respectively.

At lower densities, Zangi and Roccatano also observed that MeOH molecules form small rings, primarily consisting of four or five molecules. Although the probability of forming larger-membered rings is nonzero, it dramatically decreases to nearly zero beyond five-membered rings, with a maximum probability observed for four-membered cycles. This trend can be confirmed by visualizing molecular dynamics snapshots (Fig. 2 in Nano Lett., 2016, **16**, 3142-3147),[20] where at very low densities, only four- and five-membered rings are present, while at intermediate densities, only two six-membered and one eight-membered cycles appear. The latter results are partially consistent with the Monte Carlo simulations of MeOH deposited on carbon black of Nguyen et al.,[18] where associations forming four- and five-membered rings, were disclosed (Fig. 4 in Journal of Physical Chemistry C, 2011, **115**, 16142–16149). However, the Monte Carlo simulations treated the MeOH molecules as completely rigid entities, and some details could have been lost in the simulations. This limitation could also be behind the impossibility of detecting larger MeOH cycles (six- and seven-membered rings in particular), even when the graphene surface is fully saturated.

In this regard, our results slightly deviate from such previous findings, as simulations performed with four water molecules yielded only 4 out of 100 structures leading to four-membered cycles. Instead, an open structure where the last OH bond points toward the surface, was found to be almost equally stable and more prevalent. This result contrasts notably with the simulation involving five methanol molecules, where a higher number of structures led to five-membered cycles. In the latter simulation, only a single four-membered cycle was observed, with open structures being significantly more abundant (see SI). A similar pattern could be observed for simulations with six and seven methanol molecules, where the probability of forming larger-membered rings increases, while four- and five-membered cycles do not appear at all.

The observed trend of our MD simulations is consistent with the binding energy results shown in Fig.4, where a logarithmic interaction energy is inferred as the ring size increases. Our findings indicate that if four-membered rings form, they will likely break to form larger and more stable cycles, with five- and six-membered rings being more prevalent. This conclusion could further be supported by the stability analysis presented in Section 3.4, where only five- and six-membered cycles remain unchanged over time.

### 4.2 Stereoselective adsorption: 2D chirality of MeOH clusters

After revisiting the diffusion and behavior of MeOH on graphene, a question not yet formally documented emerges, how does an enantiospecific aggregation process occur between MeOH and graphene? To address this question, it is key to define precisely what is chiral and what is not. As noted earlier, unlike the 3D chemical space, where racemic mixtures of chiral molecules tend to self-assemble into heterochiral, rather than homochiral, domains, the notion of 2D (surface) chirality is not immediately obvious and actually both heterochiral and homochiral aggregates are observed. Highly symmetric achiral molecules, water being the paradigmatic example, can afford one of the simplest 3D chiral clusters in an idealized static environment.[1] In that cluster, three water molecules establish three hydrogen bonds forming a ring. Two of the non-covalently linked hydrogen atoms are positioned out of the plane formed by the ring in the same direction, while the third points to the opposite direction. This 3D disposition has been also observed for MeOH.[9] From a strictly geometric and static point of view, this system exhibits no symmetry plane and hence, it may exist in two non-superimposable mirror-image configurations. In the case of water, it is estimated that the energy barrier for interconversion of the free hydrogens is even lower than the zero-point energy (ZPE) of the minimum energy structure (-0.04 kcal mol$^{-1}$).[28] This scenario implies that in a real dynamic system, even at 0 K, the three-water-molecule

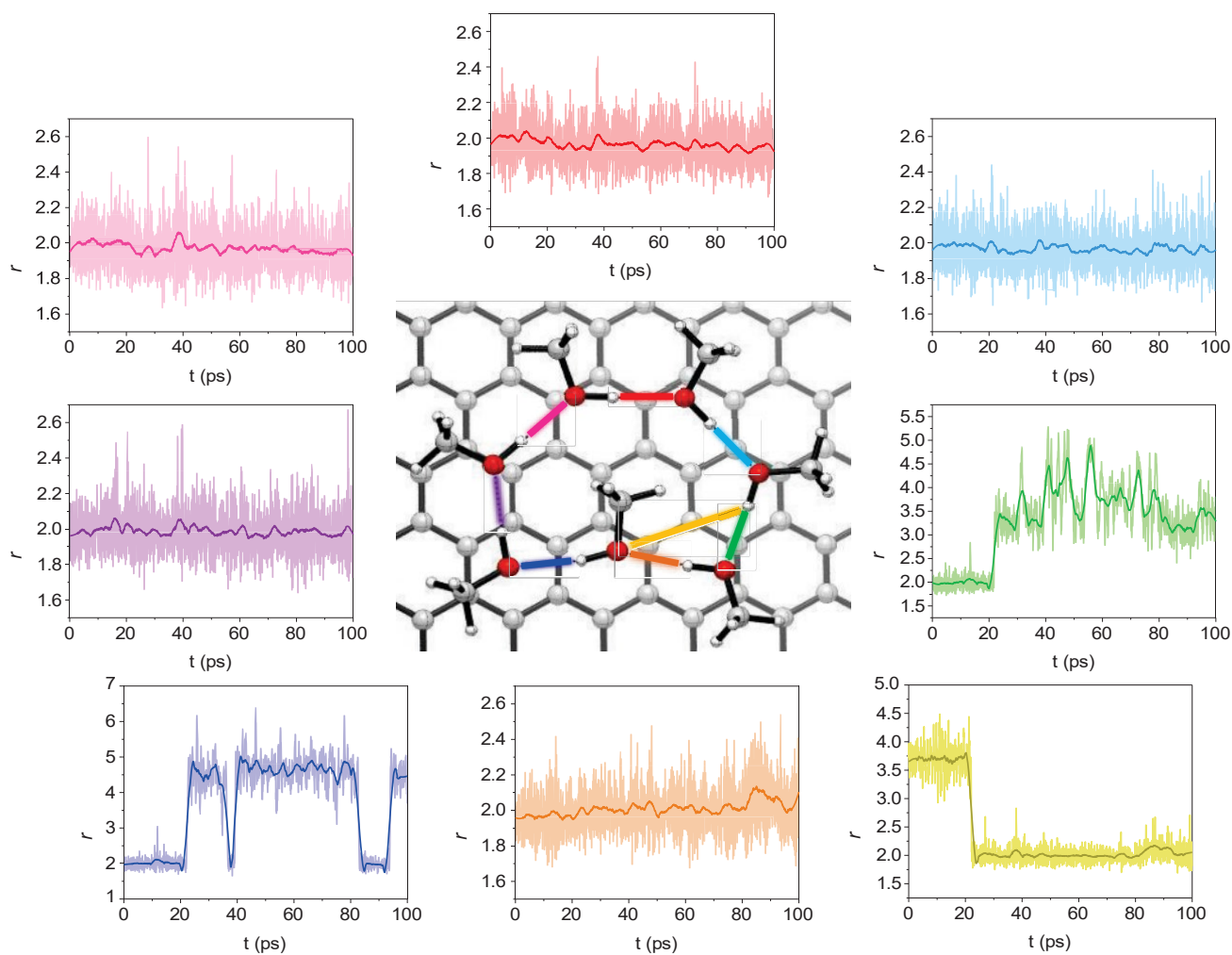

Fig. 5 MD of the seven MeOH cluster at 150 K propagated for 100 ps. The plots represent the variation of the H-bond distances (in angstroms) highlighted in the structure over the course of the MD simulation. The pale color represents the raw dynamics data, while the thinner and solid line corresponds to a Savitzky-Golay smooth with a 100-point window.

cluster remains entirely achiral due to the rapid interconversion between enantiomers.

In context and for clarity, a few representative examples of 3D, yet transient, chirality in idealized environments should be mentioned as their configurational lability may go unnoticed. This is well portrayed by non-constrained organic trisubstituted amines bearing four different substituents in almost tetrahedral geometry, which are chiral from a static point of view. However, such amines are not configurationally stable to be labeled as chiral compounds, due to the rapid and continuous interconversion of enantiomers caused by the umbrella effect at the nitrogen atom. Probably, the subtle distinction between conformation and configuration is provided by achiral alkanes. The *gauche* dispositions of butane for instance, break mirror symmetry and are therefore chiral. Nonetheless, like amines, their rapid interconversion (i.e. racemization) at ambient conditions impede to "freeze" a chiral state. In the light of the present discussion, it is pertinent to ascertain whether or not the expected racemization of the aforementioned configurationally labile species would still be feasible on a surface, where the hydrogen atoms of water, otherwise freed in the gas phase, the lone pair of amines and the chains of alkanes could now be engaged in non-covalent interactions with the surface, thereby restricting their motion.

In the case of cyclic clusters of four, five, and six-membered MeOH formed on the graphene surface at low densities, there is no possibility of racemization into clusters with the opposite configuration at working temperatures prior to evaporation, as they remain stable even at 150 K, as demonstrated by the results presented in Section 3.4. Therefore, unlike the cases mentioned above, these MeOH aggregates can be considered chiral clusters. In other words, a racemic mixture (conglomerate) is formed, different from a racemic compound or heterochiral racemate, for which each oligomer would have the two configurations in equal ratio.

Moreover, the heterochiral behavior of MeOH on graphene at high densities, results from a favorable zig-zag pattern where each oxygen atom adopts a different configuration, which renders an achiral structure. Although every methanol unit lacks the fourth non-covalent interaction that would provide it with four substituents, the HOMO molecular orbital, which points toward the graphene surface, is non-degenerate relative to HOMO-1, which is parallel to the graphene surface, and can thus be considered an additional substituent. In this context, the CIP rules would be similar to those mentioned in Section 3.2, where the lowest-priority substituent is the HOMO-1 of MeOH.

Even if spontaneous mirror-symmetry breaking is plausible, especially under kinetic control and in 3D, the existence of equal homochiral domains globally afford a racemic system. Therefore, it should be difficult to obtain an enantiopure monolayer at high MeOH concentrations. Statistically, however, at the very beginning of cluster formation, there will always be a slight excess of one type of cluster, which could influence the handedness of the subsequent MeOH molecules that will collide with it. A potential strategy to control enantioenrichment in the first monolayer would be to start the deposition at extremely low concentration and gradually adding MeOH, so that the arrangement of the initial molecules could provide a chiral bias.

**4.3 Binding energies**

The BEs obtained in our simulations rise to an asymptotic value, as a function of surface coverage. In contrast, in the experimental results carried out by Bolina et al., [16] the BEs reduce to an asymptotic value. This discrepancy may be attributed to the existence of kinked surfaces. In other words, defects and kinked edges lead to inhomogeneous layers, which are not simulated here and most likely influence the BE to a significant extent.

Much narrower BE data for submonolayers have been reported in recent studies (green solid band in Fig.4). [17] That energy range, obtained at 90 K, is also associated with H-bonded MeOH molecules rather than isolated MeOH molecules on the graphene surface. Furthermore, this BE range perfectly matches the one obtained for the first adsorption layer of MeOH on graphene-covered Pt(111) (solid orange line in Fig.4). [27] These experimental studies, together with our analysis of cluster formation and their BE, suggest that four-, five-, six-, and seven- membered rings are likely to be the most abundant species at very low concentrations of MeOH deposited on either graphite or graphene surfaces. While recent experimental data align with our calculated BE values for four- and five-membered clusters (Fig.4), six- and seven-membered rings should not be dismissed, as their BE values fall within the range of previous experiments, [16] and represent the most abundant clusters in the simulations (see ESI). The existence of significant populations of eight-to-ten-membered macrocycles is however unlikely even when falling within the range of the calculated BEs. Actually, as evidenced by rearrangements involving ten MeOH molecules, MD simulations afford more seven-membered cycles than ten-membered ones with nearly identical stability. Also, it is worth mentioning that other DFT calculations provided BEs for clusters containing three and five MeOH molecules (dashed and solid black lines in Fig.4), [14] which slightly deviate from the experimental results due to incomplete inter-molecular interactions. In fact, no cyclic oligomers were described, unlike the present simulations and those obtained using the Monte Carlo method. [18]

## 5 Conclusions

In conclusion, we have revisited and uncovered new insights into the structural organization and dynamics of cyclic and acyclic MeOH clusters adsorbed on graphene sheets, which extend and complement previous experimental and theoretical results. MD simulations show that the type of islands formed are stabilized by intermolecular H-bonding and other non-covalent interactions consisting of cyclic clusters of four, five, six, seven, and even eight MeOH molecules depending on their concentration near the graphene surface, as was concluded previously. The current analysis reveals, however, that the probability of finding cyclic five- and six-membered clusters is the highest, since four- and seven-membered clusters are not stable under typical experimental working conditions. The calculated BEs of MeOH in such structures provide a qualitative estimate that aligns with experimental results. The most significant finding is that the supramolecular arrangements of MeOH forming four-, five- and six-membered rings lead to homochiral entities by virtue of the stereogenicity at the oxygen atoms. Self-assembly takes place stereoselectively affording separate homochiral clusters, characteristic of a racemic conglomerate in two dimensions. On the other hand, seven and eight- membered clusters afford non-homochiral aggregates.

## Conflicts of interest

"There are no conflicts to declare".

## Data availability

The data supporting this article have been included as part of the Supplementary Information.

## Acknowledgements

J.G.-C and I.J.-S. acknowledge funding from grant PID2022-136814NB-I00 funded by the Spanish Ministry of Science, Innovation and Universities/State Agency of Research MICIU/AEI/ 10.13039/501100011033 and by "ERDF/EU". I.J.-S. acknowledges support from the ERC grant OPENS (project number 101125858) funded by the European Union. Views and opinions expressed are however those of the author(s) only and do not necessarily reflect those of the European Union or the European Research Council Executive Agency. Neither the European Union nor the granting authority can be held responsible for them. Computational resources were provided by CENITS and Foundation Computaex through the High-Performance Computing facility LUSITANIA-II, which are greatly appreciated. We would like express gratitude to Prof. Laurence Barron for whole-hearted discussions on molecular chirality.

## Notes and references

1 K. Liu, M. J. Elrod, J. G. Loeser, J. D. Cruzan, N. Pugliano, M. G. Brown, J. Rzepiela and R. J. Saykally, *Faraday Discuss.*, 1994, **97**, 35–41.

2 S. J. Jenkins, *Chirality at Solid Surfaces*, John Wiley & Sons, Inc., New York, 2018.

3 R. Raval, *Chirality at the Nanoscale: Nanoparticles, Surfaces, Materials, and More*, Wiley-VCH, Weinheim, 2009, pp. 191–213.

4 A. G. Mark, M. Forster and R. Raval, *ChemPhysChem*, 2011, **12**, 1474–1480.

5 K.-H. Ernst, *Surface Science*, 2013, **613**, 1–5.

6 F. Zaera, *Chem. Soc. Rev.*, 2017, **46**, 7374–7398.

7 A. J. Gellman, *Accounts of Materials Research*, 2021, **2**, 1024–1032.

8 S. Dutta and A. J. Gellman, *Chem. Soc. Rev.*, 2017, **46**, 7787–7839.

9 O. Mó, M. Yáñez and J. Elguero, *Journal of Chemical Physics*, 1997, **107**, 3592–3601.

10 J. C. Aldum, I. Jones, P. R. McGonigal, D. Spagnoli, N. D. Stapleton, G. F. Turner and S. A. Moggach, *CrystEngComm*, 2022, **24**, 7103–7108.

11 C. Červinka and G. J. O. Beran, *Chemical Science*, 2018, **9**, 4622–4629.

12 S. Kashtanov, A. Augustson, J.-E. Rubensson, J. Nordgren, H. Ågren, J.-H. Guo and Y. Luo, *Phys. Rev. B*, 2005, **71**, 104205.

13 S. L. Boyd and R. J. Boyd, *Journal of Chemical Theory and Computation*, 2007, **3**, 54–61.

14 E. Schröder, *Journal of Nanomaterials*, 2013, **2013**, 871706.

15 D. Acharya, K. Ulman and S. Narasimhan, *Journal of Physical Chemistry C*, 2021, **125**, 15012–15024.

16 A. S. Bolina, A. J. Wolff and W. A. Brown, *Journal of Chemical Physics*, 2005, **122**, 044713.

17 M. Doronin, M. Bertin, X. Michaut, L. Philippe and J. H. Fillion, *Journal of Chemical Physics*, 2015, **143**, 084703.

18 V. T. Nguyen, D. D. Do, D. Nicholson and J. Jagiello, *Journal of Physical Chemistry C*, 2011, **115**, 16142–16149.

19 T. J. Lawton, J. Carrasco, A. E. Baber, A. Michaelides and E. C. H. Sykes, *Phys. Rev. Lett.*, 2011, **107**, 256101.

20 R. Zangi and D. Roccatano, *Nano Letters*, 2016, **16**, 3142–3147.

21 S. Spicher and S. Grimme, *Angewandte Chemie International Edition*, 2020, **59**, 15665–15673.

22 C. Plett and S. Grimme, *Angewandte Chemie International Edition*, 2023, **62**, e202214477.

23 S. Spicher, C. Plett, P. Pracht, A. Hansen and S. Grimme, *Journal of Chemical Theory and Computation*, 2022, **18**, 3174–3189.

24 C. Bannwarth, S. Ehlert and S. Grimme, *Journal of Chemical Theory and Computation*, 2019, **15**, 1652–1671.

25 S. D. Feyter, P. Iavicoli and H. Xu, *Chirality at the Nanoscale: Nanoparticles, Surfaces, Materials, and More*, Wiley-VCH, Weinheim, 2009, pp. 215–245.

26 I. Alkorta, J. Elguero and P. Cintas, *Chirality*, 2015, **27**, 339–343.

27 R. S. Smith, J. Matthiesen and B. D. Kay, *Journal of Physical Chemistry A*, 2014, **118**, 8242–8250.

28 J. C. Owicki, L. L. Shipman and H. A. Scheraga, *The Journal of Physical Chemistry*, 1975, **79**, 1794–1811.